\documentstyle[12pt]{article}
\newcommand{\beq}{\begin{equation}}
\newcommand{\eeq}{\end{equation}}
\newcommand{\beqa}{\begin{eqnarray}}
\newcommand{\eeqa}{\end{eqnarray}}
\newcommand{\beqar}{\begin{eqnarray*}}
\newcommand{\eeqar}{\end{eqnarray*}}
\newcommand{\al}{\alpha}

\newcommand{\veps}{\varepsilon}

\newcommand{\ga}{\gamma}
\newcommand{\Ga}{\Gamma}

\renewcommand{\l}{\lambda}

\newcommand{\s}{\sigma}

\newcommand{\z}{\zeta}

\newcommand{\eg}{{\it e.g.,}\ }
\newcommand{\ie}{{\it i.e.,}\ }

\newcommand{\norm}[1]{\raise.3ex\hbox{:}#1\raise.3ex\hbox{:}}
\newcommand{\inn}{\!\cdot\!}
\newcommand{\bz}{\bar{z}}

\newcommand{\labell}[1]{\label{#1}} 
\newcommand{\labels}[1]{\label{#1}} 
\newcommand{\reef}[1]{(\ref{#1})}
\newcommand{\Tr}{{\rm Tr}}
\newcommand{\STr}{{\rm STr}}

\newcommand\hi{{\rm i}}

\newcommand\prt{\partial}

\newcommand\ls{\ell_s}

\newcommand\cF{{\cal F}}

\newcommand\F{{}_3F_2}
\begin{document}

\thispagestyle{empty}
\rightline{\small hep-th/0010122 \hfill McGill/00-13}
\rightline{\small\hfill IPM/P-2000/018}
\vspace*{2cm}

\begin{center}
{\bf \LARGE World-Volume Potentials}\\[.75em]
{\bf \LARGE on D-branes}
\vspace*{1cm}

Mohammad R. Garousi$^{a,b,}$\footnote{E-mail: garousi@theory.ipm.ac.ir
}
and Robert C. Myers$^{c,}$\footnote{E-mail: rcm@hep.physics.mcgill.ca}

\vspace*{0.2cm}

$^a${\it Department of Physics, Birjand University,
Birjand, Iran}\\[.5em]

$^b${\it Institute for Studies in Theoretical Physics
and Mathematics IPM}\\
{\it P.O.~Box 19395-5746, Tehran, Iran}\\[.5em]

$^c${\it Department of Physics, McGill University,
Montr\'eal, QC, H3A 2T8, Canada}\\[.5em]

\vspace{2cm} ABSTRACT
\end{center}
By evaluating string scattering amplitudes, we
investigate various low energy interactions for the massless
scalars on a nonabelian Dirichlet brane. We confirm the
existence of couplings of closed string fields to the world-volume
scalars, involving commutators of the latter.
Our results are consistent
with the recently proposed nonabelian world-volume actions
for D$p$-branes.

\vfill \setcounter{page}{0} \setcounter{footnote}{0}
\newpage

\section{Introduction}

D-branes have proven to be invaluable tools in investigating
nonperturbative properties of string theory --- see, \eg \cite{Polchin2}.
While originally conceived as simply surfaces supporting open
strings\cite{hlg}, Polchinski later elucidated
their role as dynamical
stringy solitons carrying Ramond-Ramond charges\cite{Polchin}. The low
energy action describing the dynamics of test D-branes consists
of two parts: the Born-Infeld action, which provides the kinetic
terms for the world-volume fields, and also contains the couplings of 
the D-brane to the massless Neveu-Schwarz (NS) fields in the bulk
supergravity\cite{bin}; and the Wess-Zumino action, which contains
the couplings to the massless Ramond-Ramond (RR) fields\cite{mike,cs}.
This nonlinear world-volume action 
reliably describes the physics of D-branes with surprising accuracy. 

One remarkable aspect of the D-brane story is that the $U(1)$
gauge symmetry of an individual D-brane is enhanced to a nonabelian
$U(N)$ symmetry for $N$ coincident D$p$-branes\cite{bound}. 
The form of the action for nonabelian D-branes in general background
fields was recently given in refs.~\cite{dielec,watiprep}. 
Just as for the abelian theory of an individual D-brane, the nonabelian action
describing multiple D-branes has two distinct pieces: the Born-Infeld action
\beq
{S}_{BI}=-T_p \int d^{p+1}\sigma\,\STr\left(e^{-\phi}\sqrt{-\det\left(
P\left[E_{ab}+E_{ai}(Q^{-1}-\delta)^{ij}E_{jb}\right]+
\l\,F_{ab}\right)\,\det(Q^i{}_j)}
\right),
\labell{finalbi}
\eeq
with 
\beq
E_{\mu\nu}=G_{\mu\nu}+B_{\mu\nu}
\qquad{\rm and}\qquad
Q^i{}_j\equiv\delta^i{}_j+i\lambda\,[\Phi^i,\Phi^k]\,E_{kj},
\labell{extra6}
\eeq
and the Wess-Zumino action
\beq
S_{WZ}=\mu_p\int \STr\left(P\left[e^{i\l\,\hi_\Phi \hi_\Phi} (
\sum C^{(n)}\,e^B)\right]
e^{\l\,F}\right)\ .
\labell{finalcs}
\eeq
In both of these expressions, $\lambda=2\pi \ls^2$. We refer the interested
reader to ref.~\cite{dielec} for more details on these actions
and our notation.

The displacements of the branes in the transverse space
are parameterized by the world-volume scalar fields, $\Phi^i,
\ i=p+1,\dots,9$. In the nonabelian theory, however,
the scalars are in the adjoint representation of the $U(N)$ world-volume
gauge symmetry. These scalars appear in the action in three ways: 
First, there is the explicit appearance in the first
exponential in eq.~\reef{finalcs}. Here, 
$\hi_\Phi$ denotes the interior product by $\Phi^i$ regarded
as a vector in the transverse space, {\it e.g.}, acting on an $n$-form
$C^{(n)}={1\over n!}C^{(n)}_{\mu_1\cdots\mu_n} dx^{\mu_1}\cdots dx^{\mu_n}$,
we have
\[
\hi_\Phi\hi_\Phi C^{(n)}={1\over2(n-2)!}[\Phi^i,\Phi^j]\, 
C^{(n)}_{ji\mu_3\cdots\mu_n}dx^{\mu_3}\cdots dx^{\mu_n}
\ . \labell{inside}\nonumber
\]
Next, both actions involve the pull-back (denoted by $P[\cdots]$) of various
spacetime tensors to the world-volume which now involves
covariant derivatives of the nonabelian scalars. For example, 
\[
P[E]_{ab}=E_{ab}+\l\,E_{ai}\,D_{b}\Phi^i+
\l\,E_{ib}\, D_a\Phi^i+
\l^2\,E_{ij}D_a\Phi^iD_b\Phi^j\ .
\]
Finally, the bulk supergravity fields
are in general functions of all of the spacetime coordinates, and so
in the world-volume action, they are implicitly functionals
of the nonabelian scalars.
For example, the metric functional appearing in
the D-brane action would be given by a {nonabelian} Taylor expansion
\beqa
G_{\mu\nu}&=&\exp\left[\l\Phi^i\,{\prt_{x^i}}\right]G^0_{\mu\nu}
(\sigma^a,x^i)|_{x^i=0}
\labell{slick}\nonumber\\
&=&\sum_{n=0}^\infty {\l^n\over n!}\,\Phi^{i_1}\cdots\Phi^{i_n}\,
(\prt_{x^{i_1}}\cdots\prt_{x^{i_n}})G^0_{\mu\nu}
(\sigma^a,x^i)|_{x^i=0}\ .
\nonumber
\eeqa
In both parts of the action, eqs.~\reef{finalbi} and \reef{finalcs}, there is a
single (symmetrized) gauge trace which encompasses all of the scalars
appearing in these various ways (and, of course, the gauge fields as well).

The implicit appearance of the nonabelian scalars in the functional
dependence\cite{moremike} and pull-back\cite{hull} of the background
fields was originally argued on general grounds. Both of these
suggestions can be confirmed to leading order by examining string
scattering amplitudes\cite{us}. The interesting commutator couplings to
the bulk fields
were first discovered in the construction of the nonabelian action
in refs.~\cite{dielec,watiprep}
--- see also \cite{wati1}. There, the action
was deduced by demanding that the nonabelian theory must
be consistent with T-duality. In the present paper, we confirm
the appearance of the new commutator interactions in the nonabelian
D$p$-brane action by the direct examination of string
scattering amplitudes.

The remainder of the paper is organized as follows:
In section \ref{more}, we consider the world-volume field theory
and identify an interesting set of string scattering amplitudes.
In section \ref{coupled}, we evaluate two types of amplitudes describing
the scattering of: (i) three scalars and a RR field, and (ii) two world-volume
scalars, a gauge boson and a RR field. Comparing these results with
the interactions expected in the action \reef{finalcs}, we find precise agreement.
In section \ref{morebi}, we consider a similar
set of amplitudes where the closed string field is the
Neveu-Schwarz two-form, rather than a RR field.
We conclude in section 6 with some further 
discussion of our results. In Appendix \ref{appendix},
we describe the details of evaluating the integrals necessary to calculate
the desired scattering amplitudes.


\section{Low Energy Field Theory on the World-Volume} \labels{more}

We are interested in finding evidence of various commutator
interactions appearing in the world-volume action.
Clearly the $U(1)$ components
of the world-volume fields will not participate in these interactions.
Further, any commutator will itself be in the adjoint of $SU(N)$, and so
have a vanishing trace.
To produce a nontrivial interaction then, we will need at least
three world-volume fields in the interaction. Hence the
scattering amplitudes, which we must evaluate, will involve at least
three open string states and a single closed string state. Such an 
amplitude is equivalent to a five-point open string amplitude
with unusual kinematics \cite{igor,ours}. While these amplitudes
can be evaluated (see, \eg \cite{kit}), it will be a fairly laborious
exercise. Here, we will examine the amplitudes of interest
within the world-volume field theory before proceeding
with the string scattering calculations.

To begin let us focus the discussion on the Wess-Zumino action \reef{finalcs}. 
In this part of the action, the minimal interaction of
interest will involve a bulk RR field and the two world-volume scalars entering
in the commutator. For a D$p$-brane then,
the most straightforward interactions involve the ($p$+3)-form potential
\beqa
S^{(i)}&=&i\l\mu_p\int \STr\left(P\left[\hi_\Phi \hi_\Phi
C^{(p+3)}(\s,\Phi)\right]\right)
\labell{interac}\\
&=&{i\over2}\l^2\mu_p\int d^{p+1}\s {1\over p!}(\veps^v)^{a_0\cdots a_{p}}
\left[\vphantom{1\over p+1}
\Tr\left([\Phi^j,\Phi^i]\,\prt_{a_{0}}\!\Phi^k\right)
C^{(p+3)}_{ijka_1\cdots a_{p}}(\s)
\right.\nonumber\\
&&\left.\qquad\qquad\qquad
+{1\over p+1}\Tr\left([\Phi^j,\Phi^i]\Phi^k\right)
\prt_kC^{(p+3)}_{ija_0\cdots a_{p}}(\s)+\cdots\right]\,\,\, .
\nonumber
\eeqa
In the second line, we have only explicitly kept the leading nontrivial
interactions, which involve three world-volume scalars. 
There, the third scalar arises from the pull-back in the first term
and from the nonabelian Taylor expansion in the second term.
Note with an integration by parts,
these two terms can be combined to yield the following simple result
\beq
S^{(i)}={i\over3}\l^2\mu_p\int d^{p+1}\s {1\over (p+1)!}
(\veps^v)^{a_0\cdots a_{p}}\,\Tr\left(\Phi^i\Phi^j\Phi^k\right)
F^{(p+4)}_{ijka_0\cdots a_{p}}(\s)\,\,\,,
\labell{simple}
\eeq
where $F^{(p+4)}=dC^{(p+3)}$. To verify the presence of
this interaction in the low energy effective action,
we examine the $\al'\rightarrow 0$ limit of the string amplitudes
involving three scalars and the RR ($p$+3)-form.

Generically in the limit $\al'\rightarrow 0$, we expect
string scattering amplitudes may contain massless poles
corresponding to the exchange of massless string states
arising from lower order interactions in the effective field
theory. That is, the leading low energy terms for the scalar field arise from
the Born-Infeld action \reef{finalbi}
\beq
-{\l^2T_p}\int d^{p+1}\sigma\,\Tr\left({1\over2}D^a\Phi^iD_a\Phi^i
-{1\over4}[\Phi^i,\Phi^j][\Phi^i,\Phi^j]\right)\,\,\,,
\labell{scalact}
\eeq
where, with the present conventions, $D_a\Phi^i=\prt_a\Phi^i+i[A_a,\Phi^i]$.
Hence the nonabelian scalar field theory includes the usual four-point 
interaction, and also interactions with the gauge field, \eg
the standard three-point interaction $\Tr(\prt^a\Phi^i\,[A_a,\Phi^i])$.
Hence we might expect low energy processes where two of the open
string states (in the amplitudes of interest) scatter to emit a massless
virtual particle which is absorbed by a lower order world-volume interaction
involving the RR field. Such an exchange would be
responsible for poles appearing in the string amplitude, and so a field
theory calculation must be done to properly subtract out such poles
and identify the contact interactions.

However, in fact, for the amplitude involving three scalars and the RR
($p$+3)-form, one finds there are no such contributions involving the
exchange of massless particles. Essentially, the ($p$+3)-form potential
has too many spacetime indices to produce a lower order interaction
in the ($p$+1)-dimensional world-volume theory. 
Hence we conclude, that in fact, the string amplitude will contain no massless
poles. Thus the leading contribution in the $\al'\rightarrow0$ limit will be
a set of contact terms which we should be able identify as arising from the
low energy interaction \reef{simple}.

Next we extend our analysis to consider interactions in the Wess-Zumino
action involving a bulk RR field, the two world-volume scalars entering
in the commutator, and a world-volume gauge field. For example on a
D$p$-brane, eq.~\reef{finalcs} includes
\beqa
S^{(ii)}&=&i\l^2\mu_p\int \STr\left(
P\left[\hi_\Phi \hi_\Phi C^{(p+1)}(\s,\Phi)\right] F\right)
\labell{interac2}\\
&=&{i\over4}\l^2\mu_p\int d^{p+1}\s {1\over(p-1)!}(\veps^v)^{a_0\cdots a_{p}}
\,\Tr\left([\Phi^j,\Phi^i]F_{a_0a_1}\right)
C^{(p+1)}_{ija_2\cdots a_{p}}(\s) + \cdots\,\,\, .
\nonumber
\eeqa
In fact, the analysis of the corresponding amplitude also
requires considering interactions which as above are order
$\l^2$ but arise simply from the pull-back or Taylor expansion of $C^{(p+1)}$
\beqa
S^{(iii)}
&=&{\l^2\mu_p\over2}\int d^{p+1}\s {1\over (p+1)!}(\veps^v)^{a_0\cdots a_{p}}
\left[\vphantom{p\over2}
\Tr\left(\Phi^j \Phi^i\right)
\prt_j\prt_iC^{(p+1)}_{a_0\cdots a_{p}}(\s)\right.
\nonumber\\
&&\qquad\qquad\qquad
+2(p+1)\,\Tr\left(\Phi^j D_{a_0}\Phi^i\right)
\prt_jC^{(p+1)}_{ia_1\cdots a_{p}}(\s)
\nonumber\\
&&\left.\qquad\qquad\qquad
+p(p+1)\,
\Tr\left(D_{a_0}\Phi^i\,D_{a_1}\Phi^j\right)
C^{(p+1)}_{ija_2\cdots a_{p}}(\s)
\vphantom{p\over2}\right]\,\,\, .
\labell{pare}\nonumber
\eeqa
Now again after integrating by parts a number of times,
these contributions can be rewritten as
\beqa
S^{(ii)}+S^{(iii)}&=&{\l^2\over2}\mu_p\int d^{p+1}\s {1\over (p+1)!}
(\veps^v)^{a_0\cdots a_{p}}\left[\Tr\left(\Phi^j\Phi^i\right)
\prt_jF^{(p+2)}_{ia_0\cdots a_{p}}(\s)\right.
\labell{simple2}\nonumber\\
&&\left.\qquad\qquad\qquad
+(p+1)\,\Tr\left(D_{a_0}\Phi^j\Phi^i\right)
F^{(p+2)}_{ija_1\cdots a_{p}}(\s)\right]
\nonumber\\
&=&{\l^2\over2}\mu_p\int d^{p+1}\s {1\over (p+1)!}
(\veps^v)^{a_0\cdots a_{p}}\left[\Tr\left(\Phi^j\Phi^i\right)
\prt_jF^{(p+2)}_{ia_0\cdots a_{p}}(\s)\right.
\labell{simple3}\\
&&\qquad\qquad\qquad
+(p+1)\,\Tr\left(\prt_{a_0}\Phi^j\Phi^i\right)
F^{(p+2)}_{ija_1\cdots a_{p}}(\s)
\nonumber\\
&&\left.\qquad\qquad\qquad
+2i(p+1)\,\Tr\left(A_{a_0}\Phi^j\Phi^i\right)
F^{(p+2)}_{ija_1\cdots a_{p}}(\s)\right]\,\,\, ,
\nonumber
\eeqa
where $F^{(p+2)}=dC^{(p+1)}$.
In the final expression, the last term gives the desired contact interaction
involving (a commutator of) two scalars and a single gauge field. The first
two terms are lower order in that they only involve two scalars, but these
will be relevant for determining the massless poles that appear in the
string amplitude.

In examining the latter, one might consider a virtual gluon $A$
propagating between two lower order interactions.
However, as above, one finds that there are no
interactions involving the RR ($p$+1)-form which would allow such
an exchange. On the other hand, the exchange of a virtual scalar
is possible. Here the two-scalar interactions in eq.~\reef{simple3}
allow a closed string RR potential to interact with an onshell scalar
emitting a virtual scalar. The latter is then absorbed through the
standard three-point interaction appearing in eq.~\reef{scalact} to
produce an onshell scalar and gauge field.
 
Hence we expect to find two massless poles in the string
scattering amplitude involving a bulk $C^{(p+1)}$ field, the two scalars
$\Phi$ and a world-volume gauge field $A$. To be more specific, let us
label the external states as:
\beqa
{\rm gauge\ vector:}&& A_1^a,\ k_1
\nonumber\\
{\rm transverse\ scalars:}&& \Phi_2^i,\  k_2
\nonumber\\
&&\Phi_3^j,\ k_3
\nonumber\\
{\rm RR}\ (p+1){\rm -form:}&& C_4^{p+1},\  p_4\,\,\, .
\labell{states}\nonumber
\eeqa 
For later purposes, we also define
\beq
s=-2k_1\cdot k_3\ ,
\qquad
t=-2k_1\cdot k_2\ ,
\qquad
u=-2k_2\cdot k_3\ .
\labell{stu}
\eeq
(Note that the component of $p_4$ orthogonal to the brane is not conserved
--- see, \eg ref.~\cite{igor,ours} --- and so $s+t+u=(p^\perp_4)^2$.)
With these definitions, the massless poles in the four-point amplitude 
of interest will be in the $s$ and $t$ channels.
With the field theory subtractions,  we will be able
to read off the leading contact contribution in the string
amplitude, which should match the last term in eq.~\reef{simple3}.
In fact the subtraction is simple since a quick examination of the
low energy interactions in eqs.~\reef{scalact} and \reef{simple3} shows
there can be no contributions with canceling factors of contracted momenta
 appearing in the numerator of these field theory amplitudes,
\eg $k_1\cdot k_2$ over the $s$ channel pole.
Hence the field theory subtractions correspond to subtracting the
purely pole terms out of the string scattering amplitude.

In principle, one should consider the possible appearance of pole terms of
the form $1/(p^\perp_4)^2=1/(s+t+u)$. Such contributions would arise if
there was an interaction involving the RR form and a single
world-volume field. For the case of interest, \ie a RR ($p$+1)-form
coupling to a D$p$-brane, the only such interaction in the Wess-Zumino action
\reef{finalcs} may be written as
\[
S^{(iv)}=\l\mu_p\int d^{p+1}\s {1\over (p+1)!}
(\veps^v)^{a_0\cdots a_{p}}\,\Tr\left(\Phi^i\right)\,
F^{(p+2)}_{ia_0\cdots a_{p}}(\s)\ .
\labell{onepot}\nonumber
\]
Of course, only the $U(1)$ component of the scalar contributes in this
interaction. Now in a scattering process,
the virtual scalar created by the above interaction
would have to be absorbed in an interaction involving three scalars and
a gauge field, but there are no world-volume interactions of the desired
form. Eq.~\reef{scalact} does include an interaction with two scalars
and two gauge fields, but there is no possibility to exchange a virtual
gauge field because it can not be absorbed by a RR ($p$+1)-form.
Thus the amplitude of interest should not include any contributions
proportional to $1/(s+t+u)$.

Finally we turn our attention to the nonabelian Born-Infeld action \reef{finalbi}.
Nontrivial commutator interactions appear here through the $Q$ matrix
\reef{extra6}, but as above any nontrivial interaction involving a single
commutator must also include at least one other open string field. 
The simplest case to consider is interactions involving three scalars
and the NS fields. After some calculations similar to those above,
one finds the following two interactions
\beq
S^{(v)}=i\l^2T_p\int d^{p+1}\s\left[{1\over3}\Tr(\Phi^i\Phi^j\Phi^k)
H_{ijk}+\Tr(D^a\Phi^j[\Phi^j,\Phi^i])B_{ia}\right]
\labell{sbppp}\,\,\,,
\eeq
where $B$ is the NS two-form and $H=dB$.
The first term here  arises from the expansion of the $\det(Q^i{}_j)$
factor, while the second term from the pull-back in the first determinant
factor. Note that the first determinant factor in eq.~\reef{finalbi}
contains other interactions involving the graviton or dilaton
and three scalars,
but they are higher order in $\l$, \ie they contain three derivatives.
It is also straightforward to show that the field theory
predicts there are no massless
poles in the corresponding string scattering amplitudes.

\section{Ramond-Ramond String Amplitudes} \labels{coupled}

The amplitude for a RR closed string state scattering with three
open strings consisting of either
three scalars or one gauge and two scalars on D$p$-brane
is given by
\beqa
A_{123}&=&-\frac{\l^2\mu_p}{2\sqrt{2}\pi}
\Tr\int\,dx_1dx_2dx_3d^2z_4\langle V_1^{NS}V_2^{NS}
V_3^{NS}V_4^{R-R}\rangle \,\,\, ,
\labell{a123}
\eeqa
where
\beqa
V_1^{NS}(k_1,\z_1,x_1)&=&\z_{1\mu}:V^{\mu}_{-1}(2k_1,x_1): \\ \nonumber
V_2^{NS}(k_2,\z_2,x_2)&=&\z_{2j}:V^j_0(2k_2,x_2): \\ \nonumber
V_3^{NS}(k_3,\z_3,x_3)&=&\z_{3i}:V^i_0(2k_3,x_3): \\ \nonumber
V_4^{R-R}(p_4,c_4,z_4,\bar{z}_4)&=&(P_-\Ga_{4(n)}M_p)^{AB}:V_{-\frac{1}{2}A}
(p_4,z_4):\ :V_{-\frac{1}{2}B}(p_4\cdot D,\bar{z}_4): \nonumber\,\,\, ,
\eeqa
and
\beqa
V^{\mu}_0(p,z)&=&(\prt X^{\mu}+ip\cdot \psi\psi^{\mu})e^{ip\cdot X}
\nonumber\\
V^{\mu}_{-1}(p,z)&=&e^{-\sigma}\psi^{\mu}e^{ip\cdot X}
\labell{componv}\\
V_{-\frac{1}{2}A}(p,z)&=&e^{-\frac{1}{2}\sigma}S_Ae^{ip\cdot X}\ . \nonumber
\eeqa
The vertex operators above are chosen such that they saturate the background superghost charge on the world-sheet, \ie $Q_{\sigma}=2$. In the first vertex
operator, the index $\mu$ will run over the world-volume (transverse)
directions when it represents a world-volume vector (transverse scalar)
state. Here we are using the notation of ref.~\cite{ours}. 
In particular, we have used the doubling trick \cite{igor,ours} to convert the
disk amplitude to a calculation involving only the standard holomorphic
correlators 
\beqa
\langle\, X^{\mu}(z)\,X^{\nu}(w)\,\rangle &=&-\eta^{\mu\nu}\,{\rm log}(z-w)
\nonumber\\
\langle\, \psi^{\mu}(z)\,\psi^{\nu}(w)\,\rangle
&=&-\frac{\eta^{\mu\nu}}{z-w}
\labell{propagator} \\
\langle\, \sigma(z)\,\sigma(w)\,\rangle &=&-{\rm log}(z-w) \, \, . 
\nonumber
\eeqa
The necessary correlation functions between the world-sheet fermions
and the spin operators appearing in \reef{a123} are
\beq
\langle\, :\!\psi^{\mu}(x_1)\!:\ :\!S_A(z_4)\!:\ :\!S_B(\bar{z}_4)\!:\,\rangle
\,=\,\frac{1}{\sqrt{2}}
(\ga^{\mu})^{AB}x_{14}^{-1/2}x_{15}^{-1/2}x_{45}^{-3/4}\,\,\,,
\labell{core1}
\eeq
and
\beq
:\!\psi^{\mu}\psi^{\nu}(x_2)\!:\ :\!S_A(z_4)\!: \simeq
-\frac{1}{2}(\Ga^{\mu\nu})_A{}^BS_B(z_4)x_{24}^{-1}\,\,\,,
\labell{core2}
\eeq
where $\Ga^{\mu\nu}=(\ga^{\mu}\ga^{\nu}-\ga^{\nu}\ga^{\mu})/2$ and we have
defined $x_4\equiv z_4$ ,  $x_5\equiv
\bar{z}_4$ and $x_{ij} =x_i-x_j$. Using the world-sheet fermion correlations
\reef{propagator} and \reef{core2}, one can reduce all the correlators in \reef{a123}
between the world-sheet fermions and the spin operators to the correlation function
\reef{core1}. After also performing the world-sheet bosonic correlation functions, the
scattering amplitude in eq.~\reef{a123}
can be put in the following form:
\beqa
A_{123}(C^{(n-1)},A,2\Phi)&=&\frac{\l^2\mu_p}{4\pi}{\rm
{Tr}}(\z_{1a}\z_{2j}\z_{3i})
(P_-\Ga_{4(n)}M_p)^{AB}\int\,
dx_1dx_2dx_3dx_4dx_5\,I \labell{a123one}\\
&&\left[\eta^{ij}(1-4k_2\cdot k_3)(\ga^a)_{AB}a_1+\left(p_4^ip_4^j
(\ga^a)_{AB}\right.\right.
\left.- p_4^j(\ga^ik_3\inn\ga\ga^a)_{AB} \right.\nonumber\\
&&\left.-p_4^i(\ga^jk_2\inn\ga\ga^a)_{AB}+(k_2\inn\ga\ga^j\ga^i
\ga^ak_3\inn\ga)_{AB}\right)a_2
\left.+2k_{3}^ap_4^j(\ga^i)_{AB}a_3 \right. \nonumber\\
&&\left.+\left(2k^a_2p_4^i(\ga^j)_{AB}+2k_2^a(k_3\inn\ga\ga^i\ga^j)_{AB}
\right)a_4\right.
\left.+2k_3^a(k_2\inn\ga\ga^j\ga^i)_{AB}a_5 \right. \nonumber\\
&&\left.-\left(2k_2\cdot k_3(\ga^i\ga^j\ga^a)_{AB}-2\eta^{ij}
(k_2\inn\ga k_3\inn\ga\ga^a)_{AB}
\right)a_6\right. \nonumber\\
&&\left.-4k_3^a\eta^{ij}(k_2\inn\ga)_{AB}a_7+4k_2^a\eta^{ij}
(k_3\inn\ga)_{AB}a_8\right] \,\, 
\nonumber
\eeqa
\beqa
A_{123}(C^{(n-1)},3\Phi)&\!\!\!\!\!=\!\!\!\!\!&\frac{\l^2\mu_p}{4\pi}{\rm {Tr}}
(\z_{1i}\z_{2j}\z_{3k})(P_-\Ga_{4(n)}M_p)^{AB}\int\,
dx_1dx_2dx_3dx_4dx_5\,I \labell{a1233}\\ 
&&\left[\left(p_4^jp_4^k(\ga^i)_{AB}+p_4^k(k_2\inn\ga\ga^j\ga^i)_{AB}
+p_4^j(k_3\inn\ga\ga^k\ga^i)_{AB}-(k_3\inn\ga\ga^k\ga^i\ga^j
k_2\inn\ga)_{AB}\right)a_2
\right.\nonumber\\
&&-2\eta^{ij}p_4^k(k_2\inn\ga)_{AB}a_4+\left(2\eta^{ik}
(k_3\inn\ga\ga^jk_2\inn\ga)_{AB}
-2\eta^{ik}p_4^j(k_3\inn\ga)_{AB}\right)a_{9}\nonumber\\
&&-2\eta^{ij}(k_3\inn\ga\ga^k k_2\inn\ga)_{AB}a_{10}+
\left(2\eta^{jk}(k_3\inn\ga\ga^i k_2\inn\ga)_{AB}-2k_2\inn
k_3(\ga^j\ga^k\ga^i)_{AB}\right)a_{11}
\nonumber\\
&&\left.+4k_2\inn k_3\left(-\eta^{jk}(\ga^i)_{AB}a_1+\eta^{ij}(\ga^k)_{AB}a_8-
\eta^{ik}(\ga^j)_{AB}a_{12}\right)+\eta^{jk}(\ga^i)_{AB}a_1\right]\,\,\, ,
\nonumber
\eeqa
where
\beqa
I&\equiv&x_{12}^{4k_1\cdot k_2}x_{13}^{4k_1\cdot k_3}x_{14}^{2k_1\cdot p_4}
x_{15}^{2k_1\cdot p_4}x_{32}^{4k_2\cdot k_3}
x_{24}^{2k_2\cdot p_4}x_{25}^{2k_2\cdot p_4}x_{34}^{2k_3\cdot p_4}
x_{35}^{2k_3\cdot p_4}x_{45}^{p_4\cdot D\cdot p_4}
\nonumber\\
a_1&\equiv&x_{32}^{-2}(x_{14}x_{15}x_{45})^{-1}\nonumber\\
a_2&\equiv&x_{45}(x_{34}x_{35}x_{24}x_{25}x_{14}x_{15})^{-1}\nonumber\\
a_3&\equiv&(x_{14}x_{35}x_{13}x_{24}x_{25})^{-1}\nonumber\\
a_4&\equiv&(x_{14}x_{25}x_{12}x_{34}x_{35})^{-1}\nonumber\\
a_5&\equiv&(x_{15}x_{13}x_{34}x_{24}x_{25})^{-1}\nonumber\\
a_6&\equiv&(x_{14}x_{15}x_{34}x_{25}x_{32})^{-1}\labell{a18}\\
a_7&\equiv&(x_{14}x_{45}x_{13}x_{32}x_{25})^{-1}\nonumber\\
a_8&\equiv&(x_{14}x_{45}x_{35}x_{12}x_{32})^{-1}\nonumber\\
a_{9}&\equiv&(x_{13}x_{14}x_{24}x_{25}x_{35})^{-1}\nonumber\\
a_{10}&\equiv&(x_{12}x_{24}x_{15}x_{34}x_{35})^{-1}\nonumber\\
a_{11}&\equiv&(x_{14}x_{15}x_{32}x_{24}x_{35})^{-1}\nonumber\\
a_{12}&\equiv&x_{34}(x_{13}x_{14}x_{32}x_{24}x_{35}x_{45})^{-1}\nonumber\,\,\, .
\eeqa
Note that the gamma matrices appear inside traces. That is
\beqa
(P_-\Gamma_{4(n)}M_p)^{AB}(\ga^{\mu_1}\cdots\ga^{\mu_n})_{AB}
&=&{\rm {tr}}(P_-\Gamma_{4(n)}M_p\ga^{\mu_n}\cdots\ga^{\mu_1})\ ,
\labell{nonet}
\eeqa
where we use tr($\ldots$) to denote the trace on gamma matrix indices
(as opposed to Tr($\ldots$) for the nonabelian gauge trace).

A highly nontrivial check of the results in eqs.~\reef{a123one}
and \reef{a1233} is that the integrals are $SL(2,R)$ invariant.
To remove the associated divergence and properly evaluate the amplitude,
we fix: $x_4=i$, $x_5=-i$, $x_1=R\rightarrow \infty$.
With this choice, one finds
\beqa
L_j&\equiv&\int dx_1dx_2dx_3dx_4dx_5 Ia_j\longrightarrow 
\int_{-\infty}^{\infty}dx_2\int_{x_2}^{\infty}dx_3J_j\,\,\, ,
\labell{integra}
\eeqa
where $j=1,2,\ldots,12$ and
\beqa
J_j&=&(2i)^{p_4\cdot D\cdot p_4+n_{45}^j}(x_2-i)^{2k_2\cdot p_4+n_{24}^j}
(x_2+i)^{2k_2\cdot p_4+n_{25}^j} \\ \nonumber
&&(x_3-i)^{2k_3\cdot p_4+n_{34}^j}(x_3+i)^{2k_3\cdot p_4+n_{35}^j}
(x_3-x_2)^{4k_2\cdot k_3+n_{32}^j}  \,\,\, ,
\eeqa
and integer $n_{kl}^j$ above is defined to be the power of $x_{kl}$ in $a_j$.
For example $n_{32}^1=-2$, $n_{14}^1=n_{15}^1=n_{45}^1=-1$,
$n_{24}^1=n_{25}^1=n_{34}^1=n_{35}^1=0$.
These integrals can be evaluated --- see the Appendix for details
--- and the result may be written in the following form:
\beqa
L_j&=&-(-i)^{2(t+s+u)}\Gamma(-n_{32}^j-n_{24}^j-n_{25}^j-n_{34}^j-n_{35}^j
-2-2t-2s-2u)\nonumber
\nonumber\\
&&\left(\exp[-i\pi(n_{24}^j+n_{34}^j+u)]
\sin[\pi(n_{32}^j+n_{25}^j+n_{35}^j+t+s)]\right.\nonumber\\
&&\left.\times\frac{\Gamma(1+n_{32}^j-2u)\Gamma(-1-n_{32}^j-n_{35}^j-s+u)
\Gamma(2+n_{32}^j+n_{25}^j+n_{35}^j+t+s)}{\Gamma(-n_{35}^j-s-u)
\Gamma(-n_{24}^j-n_{34}^j-t-s-2u)}\right.\nonumber\\
&&\times \F(-n_{34}^j-s-u,1+n_{32}^j-2u,2+n_{32}^j+n_{25}^j+n_{35}^j+t+s;
\nonumber\\
&&\qquad\qquad\qquad\qquad 2+n_{32}^j+n_{35}^j+s-u,-n_{24}^j-n_{34}^j-t-s-2u;1)
\labell{integ}\\
&&+\exp[-i\pi(n_{32}^j+n_{24}^j+n_{34}^j+n_{35}^j+s)]\sin[\pi(n_{25}^j+t+u)]
\nonumber\\
&&\times\frac{\Gamma(-1-n_{32}^j-n_{34}^j-n_{35}^j-2s)
\Gamma(1+n_{32}^j+n_{35}^j+s-u)\Gamma(1+n_{25}^j+t+u)}
{\Gamma(-n_{34}^j-s-u)\Gamma(-1-n_{32}^j-n_{24}^j-n_{34}^j-n_{35}^j-t-2s-u)}
\nonumber\\
&&\times \F(-n_{35}^j-s-u,-1-n_{32}^j-n_{34}^j-n_{35}^j-2s,1+n_{25}^j+t+u;
\nonumber\\
&&\left.\qquad\qquad\qquad-n_{32}^j-n_{35}^j-s+u,
-1-n_{32}^j-n_{24}^j-n_{34}^j-n_{35}^j-t-2s-u;1)\frac{}{}\right)\,\,\, ,
\nonumber
\eeqa
where $\F$ is a generalized hypergeometric function (see Appendix)
and the three independent Mandelstam variables are defined in eq.~\reef{stu}.
A careful examination of these results reveals that the
amplitudes of interest potentially have poles at
$s,t,u,s+t+u=0,\frac{1}{2},1,\frac{3}{2},2,\ldots$ reflecting the infinite tower of open string states corresponding to excitation of the D$p$-branes.

Now we are interested in the low energy limit: $s,t,u\rightarrow
0$.\footnote{We are using conventions where $\alpha'$ is fixed, \ie
$\alpha'=2$. Note that this convention fixes $\l=4\pi$.} Using the standard expansion for the gamma function, and the
following expansion for the hypergeometric function \cite{kit}
\beqa
\F(a,b,c;d,e;1)&=&1-\frac{abc}{de(a+b+c-d-e)}\left(1+[\z(2)+(a+b+c-d-e)\z(3)]
\right.\nonumber\\
&&\times [(a+b+c-d-e)(b+c-d-e)-(b-d)(c-e+b)-c(c-e)]\nonumber\\
&&+\z(3)[a(b-d)(c-e)-bd(b-d)-ce(c-e)+(a+b+c-d-e)\nonumber\\
&&\left.\times (-e(c-e)-d(b-d))]+\cdots\right)\ ,
\labell{Fabcde}\nonumber
\eeqa
one finds that, in the low energy limit, eq.~\reef{integ} yields:
\beqa
L_1^{\rm {low}}&=&-i\frac{\pi^2}{2}(u+\frac{st}{t+s+u})\nonumber \\ \nonumber
L_2^{\rm {low}}&=&i\pi^2 \\ \nonumber
L_3^{\rm {low}}&=&-\frac{\pi}{2}(\frac{1}{s})-i\frac{\pi^2}{2} \\ \nonumber
L_4^{\rm {low}}&=&\frac{\pi}{2}(\frac{1}{t})-i\frac{\pi^2}{2} \\ 
L_5^{\rm {low}}&=&-\frac{\pi}{2}(\frac{1}{s})+i\frac{\pi^2}{2}\labell{inlow} \\ \nonumber
L_6^{\rm {low}}&=&-\frac{\pi}{2}(\frac{1}{u})-i\frac{\pi^2}{2} \\ \nonumber
L_7^{\rm {low}}&=&\frac{\pi}{4}(\frac{1}{s}+\frac{1}{u})+i\frac{\pi^2}{4}
(1-\frac{t}{t+s+u}) \\ \nonumber
L_8^{\rm {low}}&=&\frac{\pi}{4}(\frac{1}{t}+\frac{1}{u})-i\frac{\pi^2}{4}
(1-\frac{s}{t+s+u}) \\ 
L_{11}^{\rm {low}}&=&-\frac{\pi}{2}(\frac{1}{u})-i\frac{\pi^2}{2}\,\, .
\nonumber
\eeqa
We have not listed the results for $L_9^{\rm low}$, $L_{10}^{\rm low}$ and $L_{12}^{\rm low}$
as they are not
needed in evaluating the amplitudes of interest in this section --- see below.

Consider the amplitude \reef{a1233} describing the scattering of the RR
field with three world-volume scalars. For the case of interest, we have
$n=p+4$ and one finds that many of the terms in eq.~\reef{a1233} vanish.
These vanishings result because after performing the trace over the gamma
matrices \reef{nonet}, some of the indices of the RR field strength (implicit
in $\Gamma_{4(n)}$) are contracted. The only non-zero
terms are
\beqa
A_{123}(C^{(p+3)},3\Phi)&=&-\frac{\l^2\mu_p}{4\pi}
{\rm {Tr}}(\z_{1i}\z_{2j}\z_{3k})(P_-\Ga_{4(p+4)}M_p)^{AB}\int\,
dx_1dx_2dx_3dx_4dx_5 \,I \nonumber\\
&&\qquad\quad\left( (k_3\inn\ga\ga^k\ga^i\ga^j k_2\inn\ga)_{AB}a_2+
2k_2\inn k_3(\ga^j\ga^k\ga^i)_{AB}a_{11}\right)
\labell{a12331}\\
&\rightarrow&-\frac{\l^2\mu_p}{4\pi}
{\rm {Tr}}(\z_{1i}\z_{2j}\z_{3k})
\left({\rm tr}(P_-\Ga_{4(p+4)}M_pk_2\inn\ga\ga^j\ga^i\ga^kk_3\inn\ga)L_2
\right.\nonumber\\
&&\left.\qquad\qquad\qquad
+2k_2\inn k_3{\rm tr}(P_-\Ga_{4(p+4)}M_p\ga^i\ga^k\ga^j)L_{11}\right)\ .
\nonumber
\eeqa
Now substituting in $L^{\rm {low}}_2$ and $L^{\rm {low}}_{11}$,
the leading order amplitude becomes
\beqa
A_{123}^{\rm {low}}(C^{(p+3)},3\Phi)&=&-\frac{\l^2\mu_p}{8}
{\rm {Tr}}(\z_{1i}\z_{2j}\z_{3k}){\rm {tr}}(P_-\Ga_{4(p+4)}M_p\ga^i\ga^k\ga^j)\,\,\, ,
\labell{a12332}\nonumber
\eeqa
where we have used the fact that the $k_2\inn\ga$ and $k_3\inn\ga$ in the
trace of gamma matrices in the first term of \reef{a12331} will
contract only with each other.
Averaging two non-cyclic orderings of open string vertex operators and
performing the trace over the gamma matrices then yields
\beqa
A^{\rm {low}}(C^{(p+3)},3\Phi)&=&-\frac{\l^2\mu_p}{(p+1)!}
{\rm {Tr}}(\z_{1i}[\z_{2j},\z_{3k}])(\varepsilon^v)_{a_0\cdots
a_{p}}({\cal{F}}_4^{(p+4)})^{ijka_0\cdots a_{p}}\,\,\, ,
\labell{a12333}\nonumber
\eeqa
where ${\cal{F}}_4^{(p+4)}$ denotes the linearized field strength.
That is \cite{ours}
\[
({\cal{F}}_4^{(n)})^{\mu_1\cdots\mu_n}=in\,p_4^{[\mu_1}
c_4^{\mu_2\cdots\mu_n]}
\labell{strong}\nonumber
\]
where $c_4$ is the polarization tensor associated with the
RR state. It clear that this leading term is precisely that
arising from an interaction in the low energy action of the
form given in eq.~\reef{simple}.

Now we turn to the case where the RR state
scatters with two scalars and one gauge field. 
Performing the integrals in eq.~\reef{a123one} as described above and
then substituting in the ${L^{\rm {low}}_j}$ given in eq.~\reef{inlow}
yields
\beqa
A_{123}^{\rm
{low}}(C^{(n-1)},A,2\Phi)&=&A_{123}^c+\frac{A_{123}^t}{t}+
\frac{A_{123}^s}{s}+
\frac{A_{123}^u}{u}+\frac{A_{123}^{t+s+u}}{t+s+u}\,\,\, ,\nonumber
\eeqa
where
\beqa
A_{123}^c&=&\frac{i\l^2\mu_p\pi}{4}{\rm {Tr}}(\z_{1a}\z_{2j}\z_{3i})
(P_-\Gamma_{4(n)}M_p)^{AB}\left(k_2\cdot k_3\eta^{ij}(\ga^a)_{AB} \right. 
\nonumber\\
&&\qquad\qquad\qquad
+p_4^ip_4^j(\ga^a)_{AB}-p_4^j(\ga^ik_3\inn\ga\ga^a)_{AB}-p_4^i
(\ga^jk_2\inn\ga\ga^a)_{AB} \nonumber\\
&&\qquad\qquad\qquad
+(k_2\inn\ga\ga^j\ga^i\ga^ak_3\inn\ga)_{AB}-k_3^ap_4^j(\ga^i)_{AB}-k_2^a
p_4^i(\ga^j)_{AB} \nonumber\\
&&\qquad\qquad\qquad
-k_2^a(k_3\inn\ga\ga^i\ga^j)_{AB}+k_3^a(k_2\inn\ga\ga^j\ga^i)_{AB}+
k_2\cdot k_3 (\ga^i\ga^j\ga^a)_{AB}  \nonumber\\
&&\qquad\qquad\qquad\left.
-\eta^{ij}(k_2\inn\ga k_3\inn\ga\ga^a)_{AB}-k_3^a\eta^{ij}
(k_2\inn\ga)_{AB}-k_2^a\eta^{ij}(k_3\inn\ga)_{AB}\right) 
\nonumber\\
A_{123}^t&=&\frac{\l^2\mu_p}{4}{\rm {Tr}}(\z_1\cdot k_2\z_{2j}\z_{3i})
(P_-\Gamma_{4(n)}M_p)^{AB}\left(p_4^i(\ga^j)_{AB}\right. \nonumber\\
&&\left. \qquad\qquad\qquad+(k_3\inn\ga\ga^i\ga^j)_{AB}+\eta^{ij}
(k_3\inn\ga)_{AB}\right) 
\nonumber\\
A_{123}^s&=&-\frac{\l^2\mu_p}{4}{\rm {Tr}}(\z_1\cdot k_3\z_{2j}\z_{3i})
(P_-\Gamma_{4(n)}M_p)^{AB}\left(p_4^j(\ga^i)_{AB}\right.
\labell{a123s}\\
&&\left.\qquad\qquad\qquad+(k_2\inn\ga\ga^j\ga^i)_{AB}+\eta^{ij}
(k_2\inn\ga)_{AB}\right) 
\nonumber\\
A_{123}^u&=&\frac{\l^2\mu_p}{4}{\rm {Tr}}(\z_{1a}\z_{2j}\z_{3i})
(P_-\Gamma_{4(n)}M_p)^{AB}\left(k_2\cdot k_3(\ga^i\ga^j\ga^a)_{AB} \right.
\nonumber\\
&&\left.\qquad\qquad\qquad-\eta^{ij}(k_2\inn\ga k_3\inn\ga\ga^a)_{AB}
-k_3^a\eta^{ij}(k_2\inn\ga)_{AB}+k_2^a\eta^{ij}(k_3\inn\ga)_{AB} \right) 
\nonumber\\
A_{123}^{t+s+u}&=&\frac{i\l^2\mu_p\pi}{4}{\rm {Tr}}(\z_{1a}\z_2\cdot\z_3)
(P_-\Gamma_{4(n)}M_p)^{AB}\left(-\frac{1}{2}ts(\ga^a)_{AB}\right.
\nonumber\\
&&\left.\qquad\qquad\qquad+tk_3^a(k_2\inn\ga)_{AB}+sk_2^a
(k_3\inn\ga)_{AB}
\vphantom{1\over2}\right) \,\, .
\nonumber
\eeqa
This result from eq.~\reef{a123one} corresponds to just one ordering of
open string states. Of course, the full scattering
amplitude includes a sum over all non-cyclic permutations,
and so the full low energy scattering amplitude has the form
\beqa
A^{\rm {low}}&=&\frac{1}{2}\left(A_{123}^c+A_{132}^c+\frac{A_{123}^t+A_{132}^s}{t}+
\frac{A_{123}^s+
A_{132}^t}{s}+\frac{A_{123}^u+A_{132}^u}{u}+\frac{A_{123}^{s+t+u}+
A_{132}^{s+t+u}}
{s+t+u}\right) \nonumber\\
&\equiv&A^c+\frac{A^t}{t}+\frac{A^s}{s}+\frac{A^u}{u}+
\frac{A^{s+t+u}}{s+t+u} \,\,.
\labell{alow}
\eeqa

For the case of interest, $n=p+2$, \ie we are considering $C^{(p+1)}$
coupling to a D$p$-brane. After performing the gamma matrice traces
in eq.~\reef{a123s}, the coefficients in eq.~\reef{alow} can be written as
\beqa
A^c&=&0 \nonumber\\
A^u&=&-\frac{\l^2\mu_p}{p!}\,u\,\left(\frac{}{} 
\Tr(\z_{1a_0}\z_2^j\z_3^i)+\Tr(\z_{1a_0}\z_3^j\z_2^i)\right)
(\cF^{(p+2)}_4)_{ija_1\cdots a_{p}}(\veps^v)^{a_0\cdots a_{p}}
\nonumber\\
A^t&=&\frac{2\l^2\mu_p}{(p+1)!}\left(\frac{}{}
\left(\Tr(\z_1\inn k_2\,\z_2^i\,\z_3\inn p_4)-
\Tr(\z_1\inn k_2\,\z_3\inn p_4\,\z_2^i)\right)
(\cF^{(p+2)}_4)_{ia_0\cdots a_p}\right.
\nonumber\\
&&\qquad\qquad\qquad
\left.-(p+1)\Tr(\z_1\inn k_2\,[\z_2^i,\z_3^j])k_{3a_0}
(\cF^{(p+2)}_4)_{ija_1\cdots a_p}\frac{}{}\right)(\veps^v)^{a_0\cdots a_p}
\nonumber\\
A^s&=&A^t(2\longleftrightarrow 3)
\nonumber\\
A^{s+t+u}&=&0 \ .
\labell{a02}\nonumber
\eeqa

Since $A^u$ above is proportional to $u$,
the amplitude has no massless $u$-channel pole.
Hence in agreement with the field theory analysis in the
previous section, the only poles are in the $s$- and $t$-channels.
In fact, we can match these poles in eq.~\reef{alow} precisely
with those arising in the amplitude calculated in the low energy world-volume
field theory. In the field theory, the $t$-channel amplitude can
be written as
\beqa
A_t^{C_4\Phi_3\Phi_2 A_1}&=&(\tilde{V}^{C_4\Phi_3\Phi})^i_{\alpha}
(\tilde{G}^{\Phi}
)^{ij}_{\alpha\beta}
(\tilde{V}^{\Phi\Phi_2 A_1})^j_{\beta} \,\,.
\labell{At}
\eeqa
The vertices and propagator above, which can be read from eqs.~\reef{scalact}
and \reef{simple3}, are 
\beqa
(\tilde{V}^{C_4\Phi_3\Phi})^i_{\alpha}&=&-\frac{N\l^2\mu_p}{(p+1)!}
\left[(\z_{3\alpha}\inn p_4)\,
(\cF^{(p+2)}_4)_{a_0\cdots a_p}^i\right.\nonumber\\
&&\qquad\qquad\qquad\left.+(p+1)k_{3a_0}{\z}_{3\alpha}^j
(\cF^{(p+2)}_4)^i{}_{ja_1\cdots a_p}\right](\veps^v)^{a_0\cdots a_p}\nonumber\\
(\tilde{V}^{\Phi\Phi_2 A_1})^j_{\beta}&=&-2i\l^2T_p\Tr
(\z_1\inn k_2[\z_2^j,T_{\beta}])
\labell{fvertex}\\
(\tilde{G}^{\Phi})^{ij}_{\alpha\beta}&=&-\frac{i}{N\l^2T_p}\frac{\delta^{ij}
\delta_{\alpha\beta}}{q^2}\,\,\, ,
\nonumber
\eeqa
where $q=k_1+k_2$. We have also written $\z^i={\z}^i_{\alpha}T_{\alpha}$
where $T_{\alpha}$ are the $U(N)$ generators with normalization
$\Tr(T_{\alpha}T_{\beta})=N\delta_{\alpha\beta}$.
We have simplified the vertex in the first line above using the Bianchi
identity $dF^{(p+2)}=0$, which yields
\[
0=\left[(p+1)p_{4a_0}(\cF^{(p+2)}_4)_{ija_1\cdots a_p}+
p_{4i}(\cF^{(p+2)}_4)_{ja_0\cdots a_p}-p_{4j}(\cF^{(p+2)}_4)_{ia_0\cdots a_p})
\right](\veps^v)^{a_0\cdots a_p}\,\,\,.
\labell{bianc}\nonumber
\]
Substituting eq.~\reef{fvertex} into eq.~\reef{At}, one finds 
\[
A_t^{C_4\Phi_3\Phi_2 A_1}=\frac{(A^t)}{t}\,\,\,,
\labell{tchan}\nonumber
\]
and so there is precise agreement between the field theory calculation
and the $t$-channel pole in the string amplitude.
A similar calculation in $s$-channel also yields agreement,
\[
A_s^{C_4\Phi_2\Phi_3 A_1}=\frac{(A^s)}{s} \,\, .
\labell{schan}\nonumber
\]
Finally it is straightforward to show that
the contact term $A^u/u$ exactly reproduces the interaction in
the last line of eq.~\reef{simple3}.

\section{Born-Infeld String Amplitudes} \label{morebi}

Finally we consider the string scattering amplitude of one closed string NS-NS
state and three open string scalars on a D$p$-brane:
\beqa
A&=&\frac{\l^2 T_p}{\pi}\Tr\int dx_1dx_2dx_3d^2z_4\,\langle
V_1^{NS}\,V_2^{NS}\,V_3^{NS}\,V_4^{NS-NS}\rangle \,\,\,,
\labell{Ansns}
\eeqa
where the vertex operators are
\beqa
V_\ell^{NS}(k_\ell,\z_\ell,x_\ell)&=&\z_{\ell i} :V_0^i(2k_\ell,x_\ell):
\qquad\ell=1,2,3\nonumber\\
V_4^{NS-NS}(p_4,\veps_4,z_4,\bar{z}_4)
&=&(\veps_4\inn D)_{\mu\nu}\,:V_{-1}^{\mu}(p_4,z_4):\ :V_{-1}^{\nu}
(p_4\inn D,\bz_4):\labell{vertechs}\nonumber\,\,,
\eeqa
with $V_0$ and $V_{-1}$ given in eq.~\reef{componv}.
Using the propagators given in eq.~\reef{propagator}, one can calculate the
correlators to produce a rather lengthy result of the same basic form
as in eqs.~\reef{a123one}
and \reef{a1233} where the various kinematic factors come with
one or three momenta. 
In an effort to reduce the calculations (and the presentation),
consider that we are interested
in the particular set of low energy interactions given in eq.~\reef{sbppp}.
Hence we must identify the contact terms in the amplitude which have only one
momentum. These can be produced in two different ways: If the kinematic 
factor has one momentum, the corresponding integrals may yield a constant
term in the low energy limit. Alternatively, in terms with three momenta,
two momenta may be contracted to yield a factor of $s,\ t$ or $u$
while the corresponding integral produces a massless pole in the same
channel. Thus one would again be left with contact terms containing a single
momentum. 

Let us begin with the first case. The relevant contributions are
\beqa
A^{(1)}&=&\frac{i\l^2 T_p}{\pi}\,\Tr(\z_{1i}\z_{2j}\z_{3k})
\,(\veps_4\inn D)_{\mu\nu}\int dx_1dx_2
dx_3dx_4dx_5 \,I
\left((\eta^{\mu\nu}\eta^{jk}p_4^i+4\eta^{jk}\eta^{i[\mu}k_1^{\nu]})a_1
\right.\nonumber\\
&&\left.
+(\eta^{\mu\nu}\eta^{ij}p_4^k+4\eta^{ij}\eta^{k[\mu}k_3^{\nu]})a_{13}
+(\eta^{\mu\nu}\eta^{ik}p_4^j+4\eta^{ik}\eta^{j[\mu}k_2^{\nu]})a_{14}\right)\,\,\,,
\labell{Aone}
\eeqa
where $a_1$ is defined in eq.~\reef{a18}, while
$a_{13}$ and $a_{14}$ are
\beqa
a_{13}&\equiv&x_{12}^{-2}(x_{34}x_{35}x_{45})^{-1}\ ,
\nonumber\\
a_{14}&\equiv&x_{13}^{-2}(x_{24}x_{25}x_{45})^{-1}\ .
\labell{a910}
\eeqa
Now one finds that in the low energy limit, the corresponding integrals
\reef{integ} yield no constant part in $L^{\rm low}_1$ (see eq.~\reef{inlow}),
$L^{\rm low}_{13}$ or $L^{\rm low}_{14}$.
Hence, this contribution \reef{Aone} to the amplitude gives no contact
terms of the desired form, \ie with one momentum.

Now we turn to the second case. The terms of interest are
\beqa
A^{(3)}&=&-\frac{2i\l^2 T_p}{\pi}\,\Tr(\z_{1i}\z_{2j}\z_{3k})
\,(\veps_4\inn D)_{\mu\nu}\int dx_1dx_2dx_3dx_4dx_5 \,I\labell{Atwo}\nonumber\\
&&\times\left( t[(\eta^{\mu\nu}\eta^{ij}p_4^k+4\eta^{ij}\eta^{k[\mu}k_3^{\nu]})
a_{13}+\eta^{i\nu}\eta^{j\mu}p_4^ka_{11}-\eta^{i\mu}\eta^{j\nu}p_4^ka_4\right.
\nonumber\\
&&\quad
 +2\eta^{ik}\eta^{j\nu}k_3^{\mu}a_{15}+2\eta^{jk}\eta^{i\nu}k_3^{\mu}a_{16}
-2\eta^{ik}\eta^{j\mu}k_3^{\nu}a_{17}-2\eta^{jk}\eta^{i\mu}k_3^{\nu}a_8]
\nonumber\\
&&\quad
+s[(\eta^{\mu\nu}\eta^{ik}p_4^j+4\eta^{ik}\eta^{j[\mu}k_2^{\nu]})a_{14}
+\eta^{i\nu}\eta^{k\mu}p_4^ja_5
-\eta^{k\nu}\eta^{i\mu}p_4^ja_{9}
\nonumber\\
&&\quad
+2\eta^{ij}\eta^{k\nu}k_2^{\mu}a_{17}-
2\eta^{jk}\eta^{i\nu}k_2^{\mu}a_{18}
-2\eta^{ij}\eta^{k\mu}k_2^{\nu}a_{15}
+2\eta^{jk}\eta^{i\mu}k_2^{\nu}a_{19}]
\nonumber\\
&&\quad
+u[(\eta^{\mu\nu}\eta^{jk}p_4^i+4\eta^{jk}\eta^{i[\mu}k_1^{\nu]})a_1
+\eta^{j\mu}\eta{k\nu}p_4^ia_{11}-\eta^{j\nu}\eta^{k\mu}p_4^ia_6
\nonumber\\
&&\quad
\left. +2\eta^{ij}\eta^{k\nu}k_1^{\mu}a_8-2\eta^{ik}\eta^{j\nu}k_1^{\mu}
a_{19}
-2\eta^{ij}\eta^{k\mu}k_1^{\nu}a_{16}+2\eta^{ik}\eta^{j\mu}k_1^{\nu}a_{18}]
\right)\ .
\nonumber
\eeqa
Beyond the $a_i$ given in eqs.~\reef{a18} and \reef{a910}, we have defined
\beqa
a_{15}&\equiv&(x_{12}x_{13}x_{25}x_{34}x_{45})^{-1}\nonumber\\
a_{16}&\equiv&(x_{12}x_{32}x_{15}x_{34}x_{45})^{-1}\nonumber\\
a_{17}&\equiv&(x_{12}x_{13}x_{24}x_{35}x_{45})^{-1}\nonumber\\
a_{18}&\equiv&(x_{13}x_{15}x_{32}x_{24}x_{45})^{-1}\nonumber\\
a_{19}&\equiv&(x_{13}x_{14}x_{32}x_{25}x_{45})^{-1}\ .
\labell{a1118}\nonumber
\eeqa
Evaluating the integrals as in eq.~\reef{integ}, one can extract
the massless poles in each. As well as those given in eq.~\reef{inlow},
we need
\beqa
L^{\rm low}_{9}&=&-\frac{\pi}{2}(\frac{1}{s})\nonumber\\
L^{\rm low}_{10}&=&\frac{\pi}{2}(\frac{1}{t})\nonumber\\
L^{\rm low}_{15}&=&\frac{\pi}{4}(\frac{1}{t}+\frac{1}{s})\nonumber\\
L^{\rm low}_{16}&=&-\frac{\pi}{4}(\frac{1}{t}+\frac{1}{u})\labell{L1118}\nonumber\\
L^{\rm low}_{17}&=&-\frac{\pi}{4}(\frac{1}{t}+\frac{1}{s})\nonumber\\
L^{\rm low}_{18}&=&-\frac{\pi}{4}(\frac{1}{s}+\frac{1}{u})\nonumber\\
L^{\rm low}_{19}&=&\frac{\pi}{4}(\frac{1}{s}+\frac{1}{u})\ .\nonumber
\eeqa
Further $L^{\rm low}_{13}$ and $L^{\rm low}_{14}$ appear in $A^{(3)}$,
but they do not have any massless poles.

With the above results, one finds the
following contact terms with only one momentum for the NS 2-form
\beqa
A^c_{123}(B,3\Phi)&=&-2i\l^2T_p\left(\Tr(\z_1\inn\veps_4\inn\z_2\,
p_4\inn\z_3)
+\Tr(\z_2\inn\veps_4\inn k_3\,\z_3\inn\z_1)\right.\nonumber\\
&&\left.\qquad-\Tr(\z_3\inn\veps_4\inn k_2\,\z_1\inn\z_2)+
{\rm cyclic\,\, permutations \,\,of \,\,(123)}\right)\,\,\, .
\labell{Ac123}\nonumber
\eeqa
This result corresponds to one ordering of the external open
string states. Adding the two non-cyclic permutations of these states
yields
\beqa
A^c(B,3\Phi)&=&\frac{1}{2}\left(A^c_{123}(B,3\Phi)+A^c_{132}(B,3\Phi)\right)
\nonumber\\
&=& -i\l^2T_p\left(\frac{1}{3}\Tr(\z_1^i\z_2^j\z_3^k+\z_2^i\z_1^j\z_3^k)
\left(p_{4i}\veps_{4jk}+p_{4j}\veps_{4ki}+p_{4k}\veps_{4ij}\right)\right.
\nonumber\\
&&\qquad\left.+\Tr[\z_2\inn\veps_4\inn k_3\,(\z_3\inn\z_1-\z_1\inn\z_3)]+
{\rm cyclic\,\, permutations \,\,of \,\,(123)}\frac{}{}\right)\,\,\, .
\labell{Ac12389}\nonumber
\eeqa
It is not difficult to verify that these contributions to the amplitude are
reproduced by the interactions in eq.~\reef{sbppp}.

Further one can show that the corresponding contact terms vanish if
one chooses to evaluate the amplitude for either a graviton or dilaton
polarization tensor. Again this is in agreement with the low energy
field theory where no interactions with a single derivative were found
for these fields. Finally one would like to verify that the string theory
amplitude has no massless poles in accord with the field theory analysis.
While calculating the entire string amplitude would be a very lengthy task,
a simple way to verify the absence of any poles is to calculate the amplitude
\reef{Ansns} for $k_1=k_2=k_3=p_4=0$. In this case, it is easy to check that
the amplitude is zero, so the whole amplitude has no massless poles.

\section{Discussion} \labels{discuss}

In the nonabelian world-volume theory of N coincident D$p$-branes,
the transverse scalars transform in the adjoint of the U(N) gauge
symmetry. Hence one has the possibility of new interactions involving commutators
$[\Phi^i,\Phi^j]$ which could not appear in the abelian theory describing a single D-brane.
The simplest example, of course, is that at leading order in the
low energy expansion, the scalars have a nontrivial potential which is
the square of two such commutators. The latter \cite{dielec} arises from
the expansion of the det($Q$) factor in the nonabelian Born-Infeld action
\reef{finalbi}. Similarly, the interactions with the bulk supergravity
fields are modified by the appearance of commutator terms. These appear
both in the Born-Infeld term \reef{finalbi}, through the contributions
involving the matrix $Q$, and in the Wess-Zumino term \reef{finalcs},
from the exponential of $\hi_\Phi\hi_\Phi$. In refs.~\cite{dielec}
and \cite{watiprep}, the existence of the commutator terms in the
nonabelian action was deduced by demanding that the nonabelian theory must
be consistent with T-duality and that it match the well-known abelian action.

In the present paper, we have confirmed the existence of a certain
class of these new couplings by the direct examination of string scattering
amplitudes. In section 4, we extracted contact terms that correspond
to the leading order commutator interactions arising from $Q$ in
the Born-Infeld action. Note that at the order studied here, there are
already contributions from both of the determinants appearing in
eq.~\reef{finalbi}. The first term in eq.~\reef{sbppp} originates
in the factor of det($Q$), while the second term is a contribution
from the $Q^{-1}$ in the first determinant factor.
In section 3, we have extracted interactions in the Wess-Zumino
action involving the first nontrivial contribution from
the exponential factor, $\exp[i\l\,\hi_\Phi \hi_\Phi]
\simeq 1+i\l\,\hi_\Phi \hi_\Phi+\cdots$. These interactions
are given in eqs.~\reef{interac} and \reef{interac2}.
Note that beyond the commutator interactions,
the calculations in section 3 also give direct evidence of 
the appearance of gauge covariant derivatives
in the pull-back expressions \cite{hull,us}. That is, in matching
the contact terms in the amplitude involving two scalars and
one gauge field, contributions involving the gauge field commutator
in $D_a\Phi^i$ were essential in producing the final result in 
eq.~\reef{simple3}.

These new commutator interactions further enrich the diverse
array of interesting physical properties displayed by
D$p$-branes. In particular, in the Wess-Zumino term \reef{finalcs},
interactions appear involving the RR potentials with a higher form degree.
Hence in the nonabelian theory, a D$p$-brane can also couple
to the RR potentials $C^{(n)}$ with $n=p+3,p+5,\ldots$ through the additional commutator
interactions. Of course, these interactions are
reminiscent of those discussed in matrix theory \cite{matrix}.
For example, the D0-brane action includes a linear coupling to $C^{(3)}$,
the potential corresponding to D2-brane charge,
\beq
i\lambda\,\mu_0\int \Tr\, P\left[\hi_\Phi \hi_\Phi C^{(3)}\right]
=i{\l\over2}\mu_0\int dt\ \Tr \left(C_{tjk}^{(3)}(\Phi,t)\,[\Phi^k,\Phi^j]
+\lambda C^{(3)}_{ijk}(\Phi,t)\,D_t\Phi^k\,[\Phi^k,\Phi^j]
\right)\ .
\labell{magic}
\eeq
The first term on the right hand side has the form of a source
for D2-brane charge, and is essentially the interaction central
to the construction of M2-branes in matrix theory with the large N
limit \cite{matrix}. However, with finite N,
this term would vanish  upon taking the trace if $C_{tjk}^{(3)}$ was
simply  a function of the world-volume coordinate $t$ since $[\Phi^k,\Phi^j]
\in{\rm SU(N)}$. Here though, eq.~\reef{magic} yields nontrivial interactions
since the three-form components  are functionals of $\Phi^i$.
Hence, while there would be no ``monopole'' coupling to D2-brane charge,
nontrivial expectation values of the scalars can give rise to couplings
to an infinite series of higher ``multipole'' moments.

As well as allowing the D$p$-branes to act as a source for higher
RR fields, these new couplings also force the D$p$-branes to respond
to a nontrivial background RR field for which the branes
would normally be regarded as neutral. That is, with nontrivial background
fields, the commutator couplings induce new terms in the scalar potential,
and hence generically one can expect that 
new extrema will be generated for the latter.
In particular, there may be nontrivial extrema
with noncommuting expectation values of the $\Phi^i$, \eg with
$\Tr\Phi^i=0$ but $\Tr(\Phi^i)^2\ne0$. This physical response
corresponds to the external field ``polarizing'' the D$p$-branes to expand
into a noncommutative world-volume geometry \cite{fuzzball}. 
Known as the ``dielectric effect'' \cite{dielec}, 
it is a direct analog of the dielectric effect
in ordinary electromagnetism. This effect was first illustrated in
ref.~\cite{dielec} with a simple toy calculation involving
D0-branes in a background four-form field strength $F^{(4)}$. The interaction
in eq.~\reef{simple}, whose presence was confirmed by the current calculations, was
the essential coupling driving the dielectric effect in that example. It was also
noted there that the D0-branes would respond to the NS three-form
$H$ in the same way because of the first interaction appearing
eq.~\reef{sbppp}.

The dielectric effect has been found to play a role in a number of
string theory contexts. For example, the resolution of certain
singularities in the AdS/CFT correspondence has been explained
in terms of external RR and/or NS fields polarizing D3-branes \cite{joemat}
--- see also \cite{morejm}. The analogous result has also been discussed in an
M-theory framework \cite{bena}. The dielectric effect is also important
in discussing the stabilization of 
D-branes in the spacetime background corresponding to a WZW model
\cite{WZW1}, as well as AdS$_m\times S^n$ backgrounds involving RR
fields. Further, one can consider more sophisticated background
field configurations which through the dielectric effect generate more
complicated noncommutative geometries \cite{sand}. The most serious
short-coming for the toy calculations in ref.~\cite{dielec} is that
the background fields were not a consistent solution of the low
energy equations of motion.  One can find solutions with
a constant background $F^{(4)}$ in M-theory, namely the AdS$_4\times$S$^7$
and AdS$_7\times$S$^4$ backgrounds --- see, \eg \cite{duff}. In lifting
D0-branes to M-theory, they become gravitons carrying momentum
in the internal space.
Hence the expanded D2-D0 system of ref.~\cite{dielec}
is related to the ``giant gravitons'' considered in
refs.~\cite{giant,goliath,lotto}. The analog of the D2-D0 bound state in
a constant background $F^{(4)}$ corresponds to M2-branes with internal
momentum expanding into AdS$_4$ \cite{goliath}, while  that in
a constant $H$ field corresponds to the M2-branes expanding on S$^4$
\cite{giant}.

It was noted in ref.~\cite{dielec} that the new potential terms which
come into play for the dielectric effect only depend on the RR field
strength, which should be expected for the results to be invariant
under the RR gauge symmetry.
Recall that in the string scattering amplitudes in section 3,
one starts with a vertex operator written in terms of the RR
field strength. Hence the resulting contact terms are naturally
derived in terms of this field strength, and as a result are
invariant under the RR gauge transformations. However, as presented
in the Wess-Zumino action \reef{finalcs}, the interactions are naturally
written in terms of the RR potentials, and thus the RR gauge
invariance is no longer manifest. Of course, for the interactions studied here,
we have shown that the two
representations of the RR couplings agree up to total derivatives.
As seen in eq.~\reef{interac} or eqs.~(\ref{interac2}--\ref{simple3}),
the necessary integration by parts requires what appears to be a
complicated interplay between terms which have completely different
origins (\eg the interior products or the Taylor expansion or the pull-back)
in the expansion of the Wess-Zumino action.
A similar discussion applies for the NS couplings
in the Born-Infeld action \reef{finalbi}. In fact, it must be true that all of the
world-volume interactions respect the appropriate spacetime gauge
symmetries. However, given the action in eqs.~\reef{finalbi} and \reef{finalcs},
it remains an exercise to confirm these invariances on a case by case basis.
Hence from this point of view, it seems that the description of the
world-volume dynamics of D-branes is still lacking at some fundamental
level.

In this paper, we have focussed our attention on limited set of
interactions in the nonabelian world-volume action. 
Of course, one could extend our calculations to make a more extensive
survey of the interactions appearing in the low energy action, and
confirm in more detail the form of the nonabelian action given in eqs.~(\ref{finalbi})
and (\ref{finalcs}). While we restrain ourselves from a complete analysis
here, we will consider one additional example below. That is, we examine
the scattering amplitude involving one gauge field, two transverse scalars
and the RR ($p\,$--1)-form potential on a D$p$-brane. This requires
only a minor extension of the calculations already presented in
section 3, namely, we set $n=p$ in eq.~\reef{a123one}. The result
provides evidence in favor of the use of the symmetrized trace in
the Wess-Zumino action \reef{finalcs}.

If one sets $n=p$ in eq.~\reef{a123one}, the string scattering amplitude
takes the form given in eq.~\reef{alow} with
\beqa
A^c&=&-\frac{i\l^3\mu_p}{2p!}\,\Tr(\z_{1a_0}\z_2^j\z_3^i)\left(
\frac{}{}p_{4i}p_{4j}(\cF^{(p)}_4)_{a_1\cdots a_p}
\right. \nonumber\\
&&\qquad\qquad\qquad
+p\,k_{3a_1}p_{4j}(\cF^{(p)}_4)_{ia_2\cdots a_p}+
p\,k_{2a_1}p_{4i}(\cF^{(p)}_4)_{ja_2\cdots a_p}
\nonumber\\
&&\left.\qquad\qquad\qquad+p(p-1)\,k_{3a_1}k_{2a_2}
(\cF^{(p)}_4)_{ija_3\cdots a_p}\frac{}{}\right)
(\veps^v)^{a_0\cdots a_p}
+\left[\frac{}{}2\leftrightarrow 3\right]
\nonumber\\
A^u&=&A^t\,\,=\,\,A^s\,\,=\,\,0
\labell{a22}\\
A^{s+t+u}&=&-\frac{i\l^3\mu_p}{2p!}\left(-\frac{1}{2}st\,
\Tr(\z_{1a_0}\z_2\cdot\z_3)+t\,\Tr(\z_1\cdot k_3\,\z_2\cdot\z_3)k_{2a_0}
\right.\nonumber\\
&&\left.\qquad\qquad\qquad+s\,\Tr(\z_1\cdot k_2\,\z_2\cdot\z_3)k_{3a_0}
\frac{}{}\right)\,
(\cF^{(p)}_4)_{a_1\cdots a_p}(\veps^v)^{a_0\cdots a_p}
+\left[\frac{}{}2\leftrightarrow 3\right]\,\,\, .
\nonumber
\eeqa
One can easily verify that the contact terms in $A^c$ can be reproduced by the field 
interactions 
\beqa
S^{(vi)}&=&\l\mu_p\int \STr\left(
P\left[C^{(p-1)}(\s,\Phi)\right] F\right)
\nonumber\\
&=&\frac{\l^3\mu_p}{4(p-1)!}\int d^{p+1}\sigma\,(\veps^v)^{a_0\cdots a_p}
\left(\frac{}{}{\rm {STr}}(F_{a_0a_1}
\Phi^i\Phi^j)\,\prt_i\prt_jC^{(p-1)}_{a_2\cdots a_p}\right.\nonumber\\
&&\qquad\qquad\qquad
+2(p-1)\,{\rm {STr}}(F_{a_0a_1}\prt_{a_2}\Phi^i\Phi^j)\,\prt_jC^{(p-1)}_{ia_3\cdots a_p}
\nonumber\\
&&\qquad\qquad\qquad\left.+\frac{}{}(p-2)(p-1)\,{\rm {STr}}(F_{a_0a_1}\prt_{a_2}
\Phi^i\prt_{a_3}\Phi^j)\,C^{(p-1)}_{ija_4\cdots a_p}\right)
\nonumber\\
&=&\frac{\l^3\mu_p}{2p!}\int d^{p+1}\sigma \,(\veps^v)^{a_0\cdots a_p}\left(\frac{}{}
{\rm {STr}}(A_{a_0}\Phi^i\Phi^j)\,
\prt_i\prt_jF^{(p)}_{a_1\cdots a_p}
\right.\nonumber\\
&&\qquad\qquad\qquad
+2p\,{\rm {STr}}(A_{a_0}\prt_{a_1}\Phi^j\Phi^i)\,\prt_iF^{(p)}_{ja_2\cdots a_p}
\nonumber\\
&&\qquad\qquad\qquad
\left.+\frac{}{}(p-1)p\,{\rm {STr}}(A_{a_0}\prt_{a_1}\Phi^i\prt_{a_2}\Phi^j)
\,F^{(p)}_{ija_3\cdots a_p}\right)\,\,\,,
\labell{s22}
\eeqa
where the symmetric trace averages over all orderings of the three
fields enclosed in each term \cite{dielec}. In particular, in the second
term of the final expression, one has
\beq
{\rm {STr}}(A_{a_0}\prt_{a_1}\Phi^j\Phi^i)\equiv\frac{1}{2}
\Tr(A_{a_0}\prt_{a_1}\Phi^j\Phi^i+A_{a_0}\Phi^i\prt_{a_1}\Phi^j)\ .
\labell{symmm}
\eeq
For the first and third terms in the final result
in eq.~\reef{s22}, the average over noncyclic permutations is
trivial because of the index symmetries of the fields in these interactions.
Note that the symmetric averaging in the trace was essential in producing interactions
which only involve $F^{(p)}$ in the final expression.

It is straightforward to verify that the field theory will not produce any
massless poles in the $s$-, $t$- or $u$-channels, in agreement with the vanishing
of $A^s$, $A^t$ and $A^u$ in the string amplitude. However, 
since $A^{s+t+u}$ is nonvanishing in eq.~\reef{a22}, the amplitude has a pole
of the form $1/(p^\perp_4)^2=1/(s+t+u)$. Such a contribution arises in the
low energy field theory if there was an interaction involving the RR form and a single
world-volume field. For the RR ($p\,$--1)-form potential
coupling to a D$p$-form, the relevant interaction may be written as
\beq
S^{(vii)}={\l\mu_p\over p!}\int d^{p+1}\s\,
(\veps^v)^{a_0\cdots a_{p}}\,\Tr\left(A_{a_0}\right)\,
F^{(p)}_{a_1\cdots a_{p}}(\s)\ .
\labell{onepota}
\eeq
which, of course, only involves the $U(1)$ component of the gauge field.
Now field theory amplitude in $(s+t+u)$-channel is given by
\beqa
A_{s+t+u}^{C_4\Phi_3\Phi_2 A_1}&=&(\tilde{V}^{C_4A})^a_{\alpha}
\,(\tilde{G}^A)_{ab,\alpha\beta}
\,(\tilde{V}^{A\Phi_3\Phi_2 A_1})^b_{\beta} \,\,\,,
\labell{astu}
\eeqa
where the propagator is derived from the standard gauge kinetic term arising in
the expansion of the Born-Infeld term:
\[
(\tilde{G}^A)_{ab,\alpha\beta}=-\frac{i}{N\l^2 T_p}\frac{\eta_{ab}
\delta_{\alpha\beta}}{q^2}\,\,\,,
\labell{propal}\nonumber
\]
where $q=k_1+k_2+k_3$, and the first vertex is derived from the interaction given above
in eq.~\reef{onepota}:
\[
(\tilde{V}^{C_4A})^a_{\alpha}=\frac{i\l\mu_p}{p!}(\cF^{(p)}_4)_{a_1\cdots a_p}
\veps^{aa_1\cdots a_p}\Tr(T_{\alpha})\ .
\labell{vertonee}\nonumber
\]
The last vertex comes from an order $\l^4$ interaction in the Born-Infeld action
\[
-\frac{\l^4 T_p}{2}\left({\rm {STr}}(D_a\Phi_i
D_b\Phi^i F^{ac}F^b{}_c)-\frac{1}{4}{\rm {STr}}(D_a\Phi_i D^a\Phi^iF^{bc}F_{bc})\right)\,\,\,,
\labell{xxxv}\nonumber
\]
and can be written as
\beqa
(\tilde{V}^{A_1\Phi_2\Phi_3 A})^b_{\beta}&=&\frac{i\l^4 T_p}{2}\left(-\frac{1}{2}st
\Tr(\z_2\cdot\z_3
\z_1^bT_{\beta})\right.\labell{stuffx}\nonumber\\
&&+s k_3^b\Tr(\z_2\cdot\z_3\z_1\cdot k_2T_{\beta})+
+t k_2^b\Tr(\z_2\cdot\z_3\z_1\cdot k_3T_{\beta})\nonumber\\
&&\left.+\frac{1}{2}s q^b\Tr(\z_2\cdot\z_3\z_1\cdot k_2T_{\beta})
+\frac{1}{2}t q^b\Tr(\z_2\cdot\z_3\z_1\cdot k_3T_{\beta})\right)
+\left[\frac{}{}2\leftrightarrow 3\right]\,\,\,,
\nonumber
\eeqa
where we used the fact that the off-shell gauge field must
be abelian. The latter implies that ${\rm {STr}}(\cdots)$ is
equivalent to $\Tr(\cdots)$ in this term.
Replacing these vertices
and propagator into \reef{astu}, one reproduces the corresponding contribution
in the string amplitude,
\[
A^{C_4\Phi_3\Phi_2 A_1}_{s+t+u}=\frac{A^{s+t+u}}{s+t+u}\,\,\,.
\labell{lasstv}\nonumber
\]

As noted in eq.~\reef{symmm} implementing the symmetric trace has a nontrivial 
effect on these calculations. In particular, having a symmetric trace
is essential in producing the expressions in eq.~\reef{s22} which are
invariant under the RR gauge symmetry. Hence being able to match
contact terms in the string scattering amplitudes with the appropriate
field theory calculations gives a nontrivial verification
that the Wess-Zumino action \reef{finalcs} must use the symmetric
trace, at least at order $\l^3$. The general principle in constructing
the action \cite{dielec,watiprep} was consistency with T-duality.
However, the symmetric trace is not required by T-duality, rather it was chosen
to match the Matrix theory results for the linearized D0-brane couplings \cite{wati1}.
The appearance
of the maximally symmetric trace in the Matrix theory calculations seems
to be essentially a requirement of supersymmetry. 

The same symmetrized trace was suggested by Tseytlin \cite{nonab} in a discussion
of the low energy gauge theory (with trivial background fields in the bulk).
There it was shown that defining the nonabelian Born-Infeld action with
this trace matched the known superstring results for the low
energy scattering of gauge fields to fourth order in the field strengths.
However, it was later shown that the symmetrized trace requires corrections
at sixth order \cite{tilt,notyet}. These problems seem to be related to the
ambiguity between covariant derivatives and field strengths in the
nonabelian theory, in that the correction terms involve commutators
of field strengths and so could be re-expressed in terms of covariant
derivatives. It is quite probable that a fully consistent
low energy action in the nonabelian theory will require the inclusion of
interactions involving arbitrarily high derivatives of the gauge field strengths.
Some progress in 
understanding the form of this action has recently been made \cite{allon}
using ideas of noncommutative field theory \cite{ncwit}.

\section*{Acknowledgments}
Research by MRG was supported by Birjand University and IPM. 
Research by RCM was supported by NSERC of Canada and Fonds FCAR du Qu\'ebec.
RCM would like to thank the Institute for Theoretical Physics at the University
of California, Santa Barbara for hospitality at various stages of this work.
Research at the ITP, UCSB was supported by NSF Grant PHY94-07194.
RCM would also like to thank Tom Banks and Per Kraus for useful
conversations. Finally, we would like to thank Neil Constable for his
critical reading of an earlier draft of this paper.

\newpage

\appendix
\section{A Useful Integral} \labels{appendix}

In this appendix, we evaluate the double integrals that appear in
our calculations of
the scattering amplitudes, \ie eq.~\reef{integra}. 
The basic integral is
\beqa
L_j&\equiv &(2i)^f\int_{-\infty}^{+\infty}dx_2\int_{x_2}^{+\infty}dx_3\,
(x_2-i)^a(x_2+i)^b(x_3-i)^c(x_3+i)^d(x_3-x_2)^e\,\,\,,
\labell{in1}\nonumber
\eeqa
where
\beqa
a&\equiv&2k_2\cdot p_4+n_{24}^j\,=\,t+u+n_{24}^j\nonumber
\\
b&\equiv&2k_2\cdot p_4+n_{25}^j\,=\,t+u+n_{25}^j\nonumber\\
c&\equiv&2k_3\cdot p_4+n_{34}^j\,=\,s+u+n_{34}^j\nonumber\\
d&\equiv&2k_3\cdot p_4+n_{35}^j\,=\,s+u+n_{35}^j\nonumber\\
e&\equiv&4k_3\cdot k_2+n_{32}^j\,=\,-2u+n_{32}^j\nonumber\\
f&\equiv&p_4\cdot D\cdot p_4+n_{45}^j\,=\,-2s-2t-2u+n_{45}^j\,\,\,,
\labell{abcdef}
\eeqa
It is relatively straightforward to evaluate the integral over $dx_3$ leaving
\beqa
L_j&=&(2i)^f\int_{-\infty}^{+\infty}dx_2\,\left\lbrace
(x_2-i)^{a+c}(x_2+i)^{b+d+e+1}
{\Gamma(-1-d-e)\Gamma(1+e)\over\Gamma(-d)}\right.\nonumber\\
&&\qquad\qquad\qquad\qquad\times{}_2F_1\left(-c,1+e,2+d+e;\frac{x_2+i}{x_2-i}\right))
\labell{in2}\\
&&\qquad\quad+\ (x_2-i)^{a+c+d+e+1}(x_2+i)^b{\Gamma(-1-c-d-e)\Gamma(1+d+e)
\over\Gamma(-c)}
\nonumber\\
&&\qquad\qquad\qquad\qquad\left.\times
{}_2F_1\left(-d,-1-c-d-e,-d-e;\frac{x_2+i}{x_2-i}\right)\right\rbrace\,\,\, .
\nonumber
\eeqa
Using the following expansion of the hypergeometric function
\[
{}_2F_1(\alpha_1, \alpha_2,\beta;z)=\frac{\Gamma(\beta)}{\Gamma(\alpha_1)
\Gamma(\alpha_2)}
\sum_{n=0}^{\infty}\frac{\Gamma(\alpha_1+n)\Gamma(\alpha_2+n)}{\Gamma(\beta+n)}
\frac{z^n}{n!}\ ,
\labell{hyperg}\nonumber
\]
one can write eq.~\reef{in2} as
\beqa
L_j&=&(2i)^f\frac{\Gamma(1+d+e)\Gamma(-d-e)}{\Gamma(-c)\Gamma(-d)}
\nonumber\\
&&\times\sum_{n=0}^{\infty}\left\{-
\frac{\Gamma(-c+n)\Gamma(1+e+n)}{n!\Gamma(2+d+e+n)}
\int_{-\infty}^{+\infty}dx_2(x_2-i)^{a+c-n}(x_2+i)^{b+d+e+1+n}
\right.\nonumber\\
&&\left.\qquad
+\frac{\Gamma(-d+n)\Gamma(-1-c-d-e+n)}{n!\Gamma(-d-e+n)}
\int_{-\infty}^{+\infty}dx_2(x_2-i)^{a+c+d+e+1-n}(x_2+i)^{b+n}\right\}\,\,\, .
\labell{in3}\nonumber
\eeqa
In simplifying this expression, we have used 
the identity
\[
\Gamma(-1-d-e)\Gamma(2+d+e)=-\Gamma(-d-e)\Gamma(1+d+e)\ .
\]
Now the integrals over $dx_2$ have the general form
\beqa
\int_{-\infty}^{+\infty}dx(x-i)^A(x+i)^B&=&-\frac{\pi(-i)^{2A}(2i)^{2+A+B}
\Gamma(-1-A-B)}{\Gamma(-A)\Gamma(-B)}
\labell{reseau}\nonumber\\
&=&\frac{(-i)^{2A}(2i)^{2+A+B}\sin(\pi A)\Gamma(1+A)\Gamma(-1-A-B)}{\Gamma(-B)}\,\,\, .
\nonumber
\eeqa
Using this result and the following identities
\beqa
a+b+c+d+e+f&=&-3\ ,\nonumber\\
(-i)^{2(y-n)}\sin[\pi(y-n)]&=&(-i)^{2y}\sin(\pi y)\ ,\nonumber\\
\Gamma(y)\Gamma(1-y)&=&\frac{\pi}{\sin(\pi y)}\ ,
\nonumber
\eeqa
one finds
\beqa
L_j&=&-\frac{\pi\Gamma(-2-a-b-c-d-e)}{\sin[\pi(d+e)]\Gamma(-c)\Gamma(-d)}
\sum_{n=0}^{\infty}\left\{\vphantom{\Gamma\over\Gamma}
-(-i)^{2(a+c)}\sin[\pi(a+c)]\right.
\nonumber\\
&&\quad\ \times\frac{\Gamma(-c+n)\Gamma(1+e+n)\Gamma(1+a+c-n)}
{n!\Gamma(2+d+e+n)\Gamma(-1-b-d-e-n)}
\nonumber\\
&&
\quad+(-i)^{2(a+c+d+e)}\sin[\pi(a+c+d+e)\nonumber\\
&&\left.\quad\ \times\frac{\Gamma(-d+n)\Gamma(-1-c-d-e+n)\Gamma(2+a+c+d+e-n)}
{n!\Gamma(-d-e+n)\Gamma(-b-n)}\right\}\ .
\labell{in4}
\eeqa
To rewrite this result in terms of the generalized hypergeometric function
${}_3F_2$, we begin by applying the identities: 
\beqa
\frac{\Gamma(1+a+c-n)}{\Gamma(-1-b-d-e-n)}&=&-\frac{\sin[\pi(b+d+e)]
\Gamma(2+b+d+e+n)}{\sin[\pi(a+c)]\Gamma(-a-c+n)}\nonumber\\
\frac{\Gamma(2+a+c+d+e-n)}{\Gamma(-b-n)}&=&-\frac{\sin(\pi b)\Gamma(1+b+n)}
{\sin[\pi(a+c+d+e)]\Gamma(-1-a-c-d-e+n)}\ .
\nonumber
\eeqa
This allows eq.~\reef{in4} to be expressed as
\beqa
L_j&=&\frac{\pi\Gamma(-2-a-b-c-d-e)}{\sin[\pi(d+e)]\Gamma(-d)\Gamma(-c)]}
\sum_{n=0}^{\infty}\left\{\vphantom{\Gamma\over\Gamma}
-(-i)^{2(a+c)}\sin[\pi(b+d+e)]\right.\nonumber\\
&&\quad\ \times\frac{\Gamma(-c+n)\Gamma(1+e+n)\Gamma(2+b+d+e+n)}
{n!\Gamma(2+d+e+n)\Gamma(-a-c+n)}\nonumber\\
&&\quad+(-i)^{2(a+c+d+e)}\sin(\pi b)\nonumber\\
&&\left.\quad\ \times\frac{\Gamma(-d+n)\Gamma(-1-c-d-e+n)\Gamma(1+b+n)}
{n!\Gamma(-d-e+n)\Gamma(-1-a-c-d-e+n)}\right\}\,\,\,.
\labell{in5}
\eeqa
Now using the definition of ${}_3F_2$
\[
{}_3F_2(\alpha_1,\alpha_2,\alpha_3;\beta_1,\beta_2;1)=
\frac{\Gamma(\beta_1)\Gamma(\beta_2)}{\Gamma(\alpha_1)\Gamma(\alpha_2)
\Gamma(\alpha_3)}
\sum_{n=0}^{\infty}\frac{\Gamma(\alpha_1+n)\Gamma(\alpha_2+n)
\Gamma(\alpha_3+n)}{\Gamma(\beta_1+n)\Gamma(\beta_2+n)n!}
\labell{ohw}\nonumber
\]
one may write eq.~\reef{in5} as
\beqa
L_j&=&-\Gamma(-2-a-b-c-d-e)
\left\{\vphantom{\Gamma\over\Gamma}
(-i)^{2(a+c)}\sin[\pi(b+d+e)]\right.\nonumber\\
&&\qquad\times \frac{\Gamma(-1-d-e)\Gamma(1+e)\Gamma(2+b+d+e)}{\Gamma(-d)
\Gamma(-a-c)}
\nonumber\\
&&\qquad\times {}_3F_2(-c,1+e,2+b+d+e;2+d+e,-a-c;1)\nonumber\\
&&\ +(-i)^{2(a+c+d+e)}\sin(\pi b)\frac{\Gamma(1+d+e)\Gamma(1+b)\Gamma(-1-c-d-e)}
{\Gamma(-c)\Gamma(-1-a-c-d-e)}\nonumber\\
&&\left.\qquad\times {}_3F_2(-d,-1-c-d-e,1+b;-d-e,-1-a-c-d-e;1)
\vphantom{\Gamma\over\Gamma}\right\}\,\,\,,
\labell{in6}
\eeqa
where the following identities have been used
\beqa
\sin[\pi(d+e)]\,\Gamma(2+d+e)&=&\frac{\pi}{\Gamma(-1-d-e)}\nonumber\\
\sin[\pi(d+e)]\,\Gamma(-d-e)&=&-\frac{\pi}{\Gamma(1+d+e)}\ .
\nonumber
\eeqa
Given the definition of the exponents \reef{abcdef}, eq.~\reef{in6} 
reduces to the result given in eq.~\reef{integ}.

\end{document}